\documentclass{article}%
\usepackage{graphicx}
\usepackage{amsmath}
\usepackage{amsfonts}
\usepackage{amssymb}%
\providecommand{\U}[1]{\protect\rule{.1in}{.1in}}

\begin{document}

\author{Antony Valentini\\Augustus College}

\begin{center}
{\LARGE Robust predictions for the large-scale cosmological power deficit from
primordial quantum nonequilibrium}

\bigskip

\bigskip

\bigskip

\bigskip

Samuel Colin\footnote{Present address: Brazilian Centre for Research in Physics, 
Rua Dr. Xavier Sigaud 150, Urca, Rio de Janeiro-RJ, 22290-180, Brazil. 
Email address: scolin@clemson.edu} and Antony
Valentini\footnote{Corresponding author: antonyv@clemson.edu}

\textit{Department of Physics and Astronomy,}

\textit{Clemson University, Kinard Laboratory,}

\textit{Clemson, SC 29634-0978, USA.}

\bigskip

\bigskip
\end{center}

\bigskip

\bigskip

\bigskip

\bigskip

\bigskip

\bigskip

The de Broglie-Bohm pilot-wave formulation of quantum theory allows the
existence of physical states that violate the Born probability rule. Recent
work has shown that in pilot-wave field theory on expanding space relaxation
to the Born rule is suppressed for long-wavelength field modes, resulting in a
large-scale power deficit $\xi(k)$ which for a radiation-dominated expansion
is found to have an approximate inverse-tangent dependence on $k$ (assuming
that the width of the initial distribution is smaller than the width of the
initial Born-rule distribution and that the initial quantum states are
evenly-weighted superpositions of energy states)\textbf{.} In this paper we
show that the functional form of $\xi(k)$ is robust under changes in the
initial nonequilibrium distribution -- subject to the limitation of a
subquantum width -- as well as under the addition of an inflationary era at
the end of the radiation-dominated phase. In both cases the predicted deficit
$\xi(k)$ remains an inverse-tangent function of $k$. Furthermore, with the
inflationary phase the dependence of the fitting parameters on the number of
superposed pre-inflationary energy states is comparable to that found
previously. Our results indicate that, for the assumed broad class of initial
conditions, an inverse-tangent power deficit is likely to be a fairly general
and robust signature of quantum relaxation in the early universe.

\bigskip

\bigskip

\bigskip

\bigskip

\bigskip

\bigskip

\bigskip

\bigskip

\bigskip

\bigskip

\bigskip

\bigskip

\section{Introduction}

According to inflationary cosmology, the temperature anisotropies in the
cosmic microwave background (CMB) were seeded by primordial quantum
fluctuations \cite{LL00,Muk05,PU09}. Precision measurements of the CMB may
therefore provide us with tests of the quantum Born probability rule at very
early times \cite{AV07,AV08,AV10,CV13,CV15,AV15}.

The de Broglie-Bohm pilot-wave formulation of quantum theory
\cite{deB28,BV09,B52a,B52b,Holl93} provides a natural extension of the usual
quantum formalism, in which corrections to the Born rule are possible. In
pilot-wave theory, the Born rule is not fundamental but only describes a state
of statistical equilibrium
\cite{AV91a,AV91b,AV92,AV96,AV01,AV02,AV07,AV08,AV09,AV10,AVPwtMw,PV06}.

One may consider a universe that begins in a state of `quantum
nonequilibrium', with non-Born rule probabilities
\cite{AV92,AV96,AV01,AV07,AV08,AV10}. The dynamics of pilot-wave theory allows
us to evolve such a state forward in time. Extensive numerical simulations
have been performed for a free scalar field evolving on a radiation-dominated
background \cite{CV13,CV15}, with a view to applying the results to a
cosmology with a radiation-dominated pre-inflationary phase \cite{PK2007,WN08}%
. The simulations yield efficient relaxation to the equilibrium Born rule at
short (sub-Hubble) wavelengths, with a suppression or retardation of
relaxation at long (super-Hubble) wavelengths. If we assume that the initial
distribution at the beginning of pre-inflation has a width smaller than the
width of the initial Born-rule distribution, then for such initial conditions
we may expect that at the beginning of inflation there will be an anomalous
deficit in the primordial spectrum at sufficiently long wavelengths
\cite{AV07,AV08,AV10,CV13,CV15}.

Data from the \textit{Planck} satellite appear to show a large-scale power
deficit in the CMB \cite{PlanckXV-2013,Planck15-XI-PowerSpec}. The CMB
two-point angular correlation function at large scales is also smaller than
expected \cite{CHSS13}. These effects could be mere statistical fluctuations
or they could amount to genuine anomalies in the primordial spectrum --
possibly arising from new physics. Theoretical models that predict a power
deficit at large scales will help to assess the nature and significance of the
apparent anomalies in the data.

It is conceivable that the large-scale power deficit in the CMB arises from
incomplete relaxation to quantum equilibrium during a pre-inflationary era
\cite{AV07,AV08,AV10,CV13,CV15}. The lengthscale at which the deficit sets in
cannot be predicted by our model as it depends on the unknown number of
inflationary e-folds. However, we are able to predict the `shape' of the
deficit as a function of wavenumber $k$. This arises by performing simulations
of quantum relaxation on an expanding radiation-dominated background for
scalar field modes of varying wavelength -- where the extent of relaxation
broadly increases with decreasing wavelength. As a result, the primordial
spectrum is diminished by a factor $\xi(k)$ which is considerably less than
$1$ for small $k$ and which approaches $1$ (approximately) for large $k$.
Specifically, we find that $\xi(k)$ varies with $k$ approximately as an
inverse-tangent \cite{CV15}.

It would be an advantage to make predictions that are broadly independent of
the details of the putative pre-inflationary era. In this paper, we show that
our prediction of an inverse-tangent power deficit function $\xi(k)$ is robust
under: (i) changes in the initial nonequilibrium distribution, and (ii) the
addition of an inflationary period at the end of the radiation-dominated phase
(adding to the time interval over which the relaxation process is simulated).
In both cases the final result for $\xi(k)$ remains an inverse-tangent
function of $k$. With the inflationary phase, the dependence of the fitting
parameters on the number of superposed pre-inflationary energy states is also
comparable to that found previously. We then have a fairly robust prediction
for the wavelength-dependence of the power deficit, which may be compared with
data \cite{PVV15}.

\section{Quantum relaxation on expanding space}

In de Broglie-Bohm pilot-wave theory \cite{deB28,BV09,B52a,B52b,Holl93}, a
system with wave function $\psi(q,t)$ has a configuration $q(t)$ with velocity%
\begin{equation}
\frac{dq}{dt}=\frac{j}{|\psi|^{2}}\ , \label{deB}%
\end{equation}
where $j=j\left[  \psi\right]  =j(q,t)$ is the Schr\"{o}dinger current
\cite{SV08}. An ensemble of systems with the same $\psi$ can have an arbitrary
distribution $\rho(q,t)$ of configurations, where $\rho(q,t)$ necessarily
obeys the continuity equation%
\begin{equation}
\frac{\partial\rho}{\partial t}+\partial_{q}\cdot\left(  \rho\dot{q}\right)
=0\ . \label{cont}%
\end{equation}
An initial distribution $\rho(q,t_{i})=\left\vert \psi(q,t_{i})\right\vert
^{2}$ evolves into a final distribution $\rho(q,t)=\left\vert \psi
(q,t)\right\vert ^{2}$ -- the state of quantum equilibrium, for which we
obtain the Born rule and the usual quantum predictions \cite{B52a,B52b}. For a
nonequilibrium ensemble we have $\rho(q,t)\neq\left\vert \psi(q,t)\right\vert
^{2}$ and the statistical predictions generally disagree with quantum theory
\cite{AV91a,AV91b,AV92,AV07,PV06}.

In pilot-wave theory, the state of quantum equilibrium may be understood to
arise dynamically by a process of relaxation or equilibration
\cite{AV91a,AV92,AV01,VW05,EC06,TRV12,SC12,ACV14}, which may have taken place
in the early universe \cite{AV91a,AV91b,AV92,AV96,AV01,AV07,AV08,AV10}.

For our purposes, it suffices to consider a free, minimally-coupled, real
massless scalar field $\phi$ on an expanding background with line element
$d\tau^{2}=dt^{2}-a^{2}d\mathbf{x}^{2}$ where $a=a(t)$ is the scale factor
(with $c=1$). Working with Fourier components%
\begin{equation}
\phi_{\mathbf{k}}=\frac{\sqrt{V}}{(2\pi)^{3/2}}\left(  q_{\mathbf{k}%
1}+iq_{\mathbf{k}2}\right)  \ ,
\end{equation}
where $V$ is a normalisation volume and the $q_{\mathbf{k}r}$'s ($r=1,2$) are
real, the field Hamiltonian is a sum $H=\sum_{\mathbf{k}r}H_{\mathbf{k}r}$
where%
\begin{equation}
H_{\mathbf{k}r}=\frac{1}{2a^{3}}\pi_{\mathbf{k}r}^{2}+\frac{1}{2}%
ak^{2}q_{\mathbf{k}r}^{2}\ . \label{Ham}%
\end{equation}
For an unentangled mode $\mathbf{k}$ with wave function $\psi_{\mathbf{k}%
}(q_{\mathbf{k}1},q_{\mathbf{k}2},t)$, we have the Schr\"{o}dinger equation
(dropping the index $\mathbf{k}$) \cite{AV07,AV08,AV10}%
\begin{equation}
i\frac{\partial\psi}{\partial t}=\sum_{r=1,\ 2}\left(  -\frac{1}{2a^{3}%
}\partial_{r}^{2}+\frac{1}{2}ak^{2}q_{r}^{2}\right)  \psi
\end{equation}
for $\psi=\psi(q_{1},q_{2},t)$ (where $q_{1},q_{2}$ are proportional to the
real and imaginary parts of the field mode) and the de Broglie-Bohm velocities%
\begin{equation}
\dot{q}_{r}=\frac{1}{a^{3}}\operatorname{Im}\frac{\partial_{r}\psi}{\psi}
\label{deB3}%
\end{equation}
for the configuration $(q_{1},q_{2})$ (with $\partial_{r}\equiv\partial
/\partial q_{r}$). The marginal distribution $\rho=\rho(q_{1},q_{2},t)$
obeys\footnote{The marginal $\rho(q_{1},q_{2},t)$ is the probability
distribution for the field mode, which may be obtained from the complete field
probability distribution by integrating over the degrees of freedom of the
other field modes -- though such an operation is redundant if for simplicity
we assume a product distribution over modes.}%
\begin{equation}
\frac{\partial\rho}{\partial t}+\sum_{r=1,\ 2}\partial_{r}\left(  \rho\frac
{1}{a^{3}}\operatorname{Im}\frac{\partial_{r}\psi}{\psi}\right)  =0\ .
\label{CE2D}%
\end{equation}

Mathematically, a single field mode is equivalent to a two-dimensional
oscillator with time-dependent mass $m=a^{3}$ and time-dependent angular
frequency $\omega=k/a$ \cite{AV07,AV08}. This is in turn equivalent to a
standard oscillator -- with constant mass and constant angular frequency --
but with appropriately rescaled variables for each system and with standard
time $t$ replaced by a `retarded time' $t_{\mathrm{ret}}(t,k)$, where the
function $t_{\mathrm{ret}}=t_{\mathrm{ret}}(t,k)$ may be determined
\cite{CV13}.

Cosmological relaxation over a time $t$ for a field mode of wavenumber $k$ may
then be conveniently studied in terms of a standard oscillator evolving over a
retarded time $t_{\mathrm{ret}}(t,k)$. Specifically, the time evolution from
initial conditions at $t_{i}$ to final conditions at $t_{f}$ may be obtained
by evolving a standard oscillator -- with the same initial conditions -- from
the same initial time $t_{i}$ up to the retarded final time $t_{\mathrm{ret}%
}(t_{f},k)$. Note, however, that this is merely a convenient means of evolving
the continuity equation (\ref{CE2D}), which could be integrated directly
(though less efficiently) yielding the same results.

Extensive results have been obtained for a radiation-dominated expansion
($a\propto t^{1/2}$) \cite{CV13,CV15}. In the short-wavelength (sub-Hubble)
limit, $t_{\mathrm{ret}}(t,k)\rightarrow t$ and we recover the usual rapid
relaxation for a superposition of energy states (as on Minkowski spacetime).
At long (super-Hubble) wavelengths, $t_{\mathrm{ret}}(t,k)<<t$ and relaxation
is suppressed or retarded.

The suppression of relaxation at long wavelengths may be quantified in terms
of a deficit in the width of the nonequilibrium distribution. Each degree of
freedom $q_{r}$ has an equilibrium variance $\Delta_{r}^{2}=\left\langle
q_{r}^{2}\right\rangle _{\mathrm{QT}}-\left\langle q_{r}\right\rangle
_{\mathrm{QT}}^{2}$ and a nonequilibrium variance $D_{r}^{2}=\left\langle
q_{r}^{2}\right\rangle -\left\langle q_{r}\right\rangle ^{2}$, where
$\left\langle ...\right\rangle _{\mathrm{QT}}$ and $\left\langle
...\right\rangle $ are respective averages with respect to $\left\vert
\psi(q_{1},q_{2},t)\right\vert ^{2}$ and $\rho(q_{1},q_{2},t)$. The function%
\[
\xi(k)\equiv\frac{D_{1}^{2}+D_{2}^{2}}{\Delta_{1}^{2}+\Delta_{2}^{2}}%
\]
then measures the width deficit of nonequilibrium relative to equilibrium.

We have argued elsewhere \cite{AV07,AV10,CV13,CV15} that if there was a
radiation-dominated pre-inflationary era then we may expect the spectrum of
primordial perturbations generated during inflation to take the form%
\begin{equation}
\mathcal{P}_{\mathcal{R}}(k)=\mathcal{P}_{\mathcal{R}}^{\mathrm{QT}}%
(k)\xi(k)\ ,
\end{equation}
where $\mathcal{P}_{\mathcal{R}}^{\mathrm{QT}}(k)$ is the standard
quantum-theoretical spectrum. It is of particular interest to make predictions
for $\xi(k)$ with a view to comparing with CMB data.

In previous work \cite{CV13,CV15} we considered initial wave functions for a
single mode (in a putative pre-inflationary era) that are superpositions%
\begin{equation}
\psi(q_{1},q_{2},t_{i})=\frac{1}{\sqrt{M}}\sum_{n_{1}=0}^{\sqrt{M}-1}%
\sum_{n_{2}=0}^{\sqrt{M}-1}e^{i\theta_{n_{1}n_{2}}}\Phi_{n_{1}}(q_{1}%
)\Phi_{n_{2}}(q_{2}) \label{psi_i}%
\end{equation}
of $M$ energy eigenstates $\Phi_{n_{1}}\Phi_{n_{2}}$ of the initial
Hamiltonian, with coefficients $c_{n_{1}n_{2}}(t_{i})=(1/\sqrt{M}%
)e^{i\theta_{n_{1}n_{2}}}$ of equal amplitude and with randomly-chosen phases
$\theta_{n_{1}n_{2}}$. The wave function at time $t$ is%
\begin{equation}
\psi(q_{1},q_{2},t)=\frac{1}{\sqrt{M}}\sum_{n_{1}=0}^{\sqrt{M}-1}\sum
_{n_{2}=0}^{\sqrt{M}-1}e^{i\theta_{n_{1}n_{2}}}\psi_{n_{1}}(q_{1}%
,t)\psi_{n_{2}}(q_{2},t)\ , \label{psi_t}%
\end{equation}
where $\psi_{n_{r}}(q_{r},t)$ is the time evolution of $\psi_{n_{r}}%
(q_{r},t_{i})=\Phi_{n_{r}}(q_{r})$ under a formal one-dimensional
Schr\"{o}dinger equation with Hamiltonian $\hat{H}_{r}(t)$ (given by equation
(\ref{Ham})) and where the exact solution for $\psi_{n}(q,t)$ with $a\propto
t^{1/2}$ was obtained in ref. \cite{CV13}. For simplicity we assumed an
initial nonequilibrium distribution%
\begin{equation}
\rho(q_{1},q_{2},t_{i})=|\Phi_{0}(q_{1})\Phi_{0}(q_{2})|^{2}=\frac{\omega
_{i}m_{i}}{\pi}e^{-m_{i}\omega_{i}q_{1}^{2}}e^{-m_{i}\omega_{i}q_{2}^{2}}
\label{rho_i}%
\end{equation}
(equal to the equilibrium distribution for the ground state $\Phi_{0}%
(q_{1})\Phi_{0}(q_{2})$). We then performed a numerical simulation of the time
evolution $\rho(q_{1},q_{2},t)$ up to a final time $t_{f}$.

Such simulations were carried out for varying values of $k$, as well as for
varying values of $M$ and $t_{f}$. For each case we could compare the final
nonequilibrium variance with the final equilibrium variance and hence
calculate the ratio $\xi$ for varying values of $k$, $M$ and $t_{f}$. In fact,
before taking their ratio the nonequilibrium and equilibrium variances were
averaged over mixtures of initial wave functions (\ref{psi_i}) with
randomly-chosen phases. Our calculated function $\xi(k)$ -- for given $M$ and
$t_{f}$ -- is then in fact given by%
\begin{equation}
\xi(k)=\frac{\left\langle D_{1}^{2}+D_{2}^{2}\right\rangle _{\mathrm{mixed}}%
}{\left\langle \Delta_{1}^{2}+\Delta_{2}^{2}\right\rangle _{\mathrm{mixed}}%
}\ , \label{xi_mixed}%
\end{equation}
where $\left\langle ...\right\rangle _{\mathrm{mixed}}$ is an average over the
mixed ensemble of quantum states (see ref. \cite{CV15} for details).

Our numerical simulations yielded functions $\xi(k)$ that approximately take
the form \cite{CV15}%
\begin{equation}
\xi(k)=\tan^{-1}(c_{1}\frac{k}{\pi}+c_{2})-\frac{\pi}{2}+c_{3} \label{numksi}%
\end{equation}
(neglecting oscillations of amplitude $\lesssim10\%$ around the fitted curve),
where the best-fit parameters $c_{1}$, $c_{2}$, $c_{3}$ depend on $M$ and
$t_{f}$. Our results for $c_{1}$, $c_{2}$, $c_{3}$ are listed in tables I and
II of ref. \cite{CV15}.

Note that $\xi\rightarrow c_{3}$ for large $k$, where $c_{3}$ is generally
found to differ slightly from $1$. There is a nonequilibrium `residue' at
short wavelengths. For a discussion of the significance of this, see ref.
\cite{CV15}.

Our aim is to find a general signature of quantum relaxation in the early
universe that is as far as possible independent of details of the
pre-inflationary era. The `quantum relaxation spectrum' (\ref{numksi}) appears
to be a good candidate for such a signature. Here we extend the generality of
this spectrum in two ways. First, we study initial nonequilibrium
distributions that are more complicated than (\ref{rho_i}). Second, as a first
step towards a more realistic cosmological model, we include a period of
exponential expansion after the radiation-dominated phase. In both cases we
still obtain a spectral deficit $\xi(k)$ of the form (\ref{numksi}).

\section{Quantum relaxation for different initial nonequilibria}

In this section we consider cosmological quantum relaxation for three
different initial nonequilibrium distributions:%
\begin{align}
\rho_{1}(q_{1},q_{2},t_{i})  &  =|\Phi_{0}(q_{1})\Phi_{0}(q_{2})|^{2}%
\ ,\label{rho1i}\\
\rho_{2}(q_{1},q_{2},t_{i})  &  =\frac{1}{2}|\Phi_{0}(q_{1})\Phi_{0}%
(q_{2})+\Phi_{1}(q_{1})\Phi_{1}(q_{2})|^{2}\ ,\label{rho2i}\\
\rho_{3}(q_{1},q_{2},t_{i})  &  =\frac{2}{3}
|\Phi_{0}(q_{1})\Phi_{0}(q_{2})+\frac{1}{2}\Phi_{1}(q_{1})\Phi_{1}(q_{2})+\frac{1}{2}\Phi_{2}(q_{1})\Phi_{2}(q_{2})|^{2}\ ,
\label{rho3i}%
\end{align}
where as above $\Phi_{n_{1}}(q_{1})\Phi_{n_{2}}(q_{2})$ are energy eigenstates
of the initial two-dimensional harmonic oscillator Hamiltonian.

As before we take the initial wave function $\psi(q_{1},q_{2},t_{i})$ to be a
superposition (\ref{psi_i}) of $M$ energy eigenstates $\Phi_{n_{1}}(q_{1}%
)\Phi_{n_{2}}(q_{2})$ with coefficients of equal amplitude and with
randomly-chosen phases. We shall consider in particular the cases $M=9$ and
$M=25$. For these initial wave functions, the initial distributions
(\ref{rho1i})--(\ref{rho3i}) are all nonequilibrium states. The first
(\ref{rho1i}) is that used in previous simulations (equation (\ref{rho_i})):
it is equal to what would be the equilibrium distribution if the wave function
were simply the ground state. The second (\ref{rho2i}) contains an admixture
of terms from the excited state $\Phi_{1}(q_{1})\Phi_{1}(q_{2})$, while the
third (\ref{rho3i}) is equal to what would be the equilibrium distribution if
the wave function were simply $\Phi_{1}(q_{1})\Phi_{1}(q_{2})$.

Note that in pilot-wave theory the initial distribution can in principle be
any arbitrary probability function $\rho(q_{1},q_{2},t_{i})\neq|\psi
(q_{1},q_{2},t_{i})|^{2}$. The above three probability functions related to
$\Phi_{0}(q_{1})\Phi_{0}(q_{2})$ and $\Phi_{1}(q_{1})\Phi_{1}(q_{2})$ are
chosen for convenience only.

We shall consider these different initial nonequilibria in order to study two questions.

First, we study relaxation (illustrated with density plots) for the case of
$M=25$ modes. In particular, we compare results obtained with and without
spatial expansion for each of the three initial nonequilibria, thereby
extending the analysis of ref. \cite{CV13}.

Second, for the case of $M=9$ modes we perform simulations of the deficit
function $\xi(k)$ for the three different initial nonequilibria, thereby
extending the analysis of ref. \cite{CV15}.

The three initial nonequilibria (\ref{rho1i})--(\ref{rho3i}) differ
considerably (see the plots below). And yet, as we shall see, relaxation
proceeds similarly for each in the absence of spatial expansion, while in the
presence of spatial expansion there is a similar relaxation suppression for
the three cases. Crucially, the final results for the deficit function
$\xi(k)$ remain of the inverse-tangent form (\ref{numksi}), though with
coefficients $c_{1}$, $c_{2}$, $c_{3}$ that vary with the initial
nonequilibrium distribution.

\subsection{Relaxation and relaxation suppression}

In ref. \cite{CV13} we simulated relaxation on expanding space for the
illustrative case of $M=25$ modes. We compared results obtained in the absence
of spatial expansion (equivalent to the short-wavelength limit) with results
obtained in the presence of spatial expansion and in particular at long
(super-Hubble) wavelengths. For the former, we found efficient and close
relaxation to quantum equilibrium. For the latter, we found a retardation or
suppression of relaxation. These studies were carried out with the fixed
initial nonequilibrium distribution (\ref{rho1i}). Here we present comparable
simulations for all three initial densities (\ref{rho1i})--(\ref{rho3i}), with
similar results (again for $M=25$).

As in ref. \cite{CV13}, we begin at an initial time $t_{i}=10^{-4}$ and we
evolve to a final time $t_{f}=10^{-2}$ (units $\hslash=c=1$), with a scale
factor $a(t)=a_{0}(t/t_{0})^{1/2}$ where for convenience we take $a_{0}=1$ at
$t_{0}=1$. We consider a mode of wavenumber $k=10\pi$. With these values, the
initial physical wavelength $\lambda_{\mathrm{phys}}(t_{i})=a_{i}(2\pi/k)$ is
ten times the initial Hubble radius $H_{i}^{-1}=2t_{i}$ while the final
physical wavelength $\lambda_{\mathrm{phys}}(t_{f})=a_{f}(2\pi/k)$ is equal to
the final Hubble radius $H_{f}^{-1}=2t_{f}$. In other words,\ we study a field
mode that begins an order of magnitude outside the Hubble radius and we evolve
it to the time $t_{f}=t_{\mathrm{enter}}$ of Hubble entry.

We plot the evolving nonequilibrium and equilibrium distributions at the
initial and final times $t_{i}$ and $t_{f}=t_{\mathrm{enter}}$ as well as at
the intermediate time $0.5t_{\mathrm{enter}}$. As explained in ref.
\cite{CV13}, because the support of $|\psi(q_{1},q_{2},t)|^{2}$ in the
$q_{1}-q_{2}$ plane shrinks with time it is convenient to plot the
distributions in terms of appropriately rescaled variables $q_{1}^{\prime
},q_{2}^{\prime}$ with an actual density $\rho^{\prime}(q_{1}^{\prime}%
,q_{2}^{\prime},t)$ and an equilibrium density $\rho_{\mathrm{QT}}^{\prime
}(q_{1}^{\prime},q_{2}^{\prime},t)$. Furthermore, it is convenient to plot
smoothed densities $\tilde{\rho}^{\prime}(q_{1}^{\prime},q_{2}^{\prime},t)$
and $\tilde{\rho}_{\mathrm{QT}}^{\prime}(q_{1}^{\prime},q_{2}^{\prime},t)$
obtained by coarse-graining with overlapping coarse-graining cells \cite{CV13}.

The results in the absence of spatial expansion are shown in Figure 1. The
first, second and third columns show the time evolution of the initial
nonequilibrium densities (\ref{rho1i}), (\ref{rho2i}) and (\ref{rho3i})
respectively. The fourth column shows the time evolution of the quantum
equilibrium density (for a wave function with a superposition of $M=25$ energy
states). The relaxation is seen to be excellent for all three initial
nonequilibria. This answers a query that is sometimes raised: relaxation on a
fixed background occurs efficiently not only for the initial `ground-state'
nonequilibrium distribution (\ref{rho1i}) but also for other initial
nonequilibria. In the three cases studied here, the results are in fact comparable.%

\begin{figure}
[ptb]
\begin{center}
\includegraphics[width=\textwidth]
{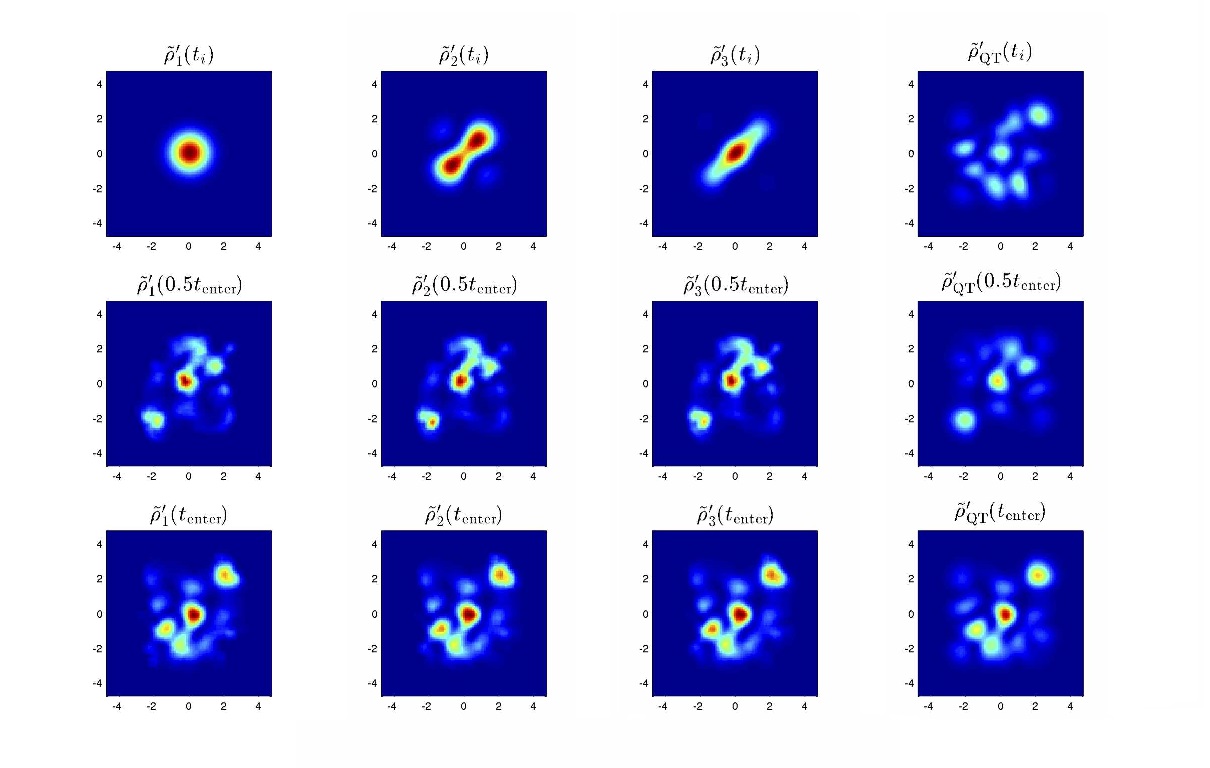}%
\caption{Relaxation in the absence of spatial expansion for different initial
nonequilibria.}%
\end{center}
\end{figure}

The results in the presence of spatial expansion are shown in Figure 2. Again,
the first, second and third columns show the time evolution of the initial
nonequilibrium densities (\ref{rho1i}), (\ref{rho2i}) and (\ref{rho3i})
respectively while the fourth column shows the time evolution of the quantum
equilibrium density (for $M=25$). As is plain from the plots, at super-Hubble
wavelengths on expanding space we find a comparable retardation or suppression
of relaxation in all three cases.%

\begin{figure}
[ptb]
\begin{center}
\includegraphics[width=\textwidth]
{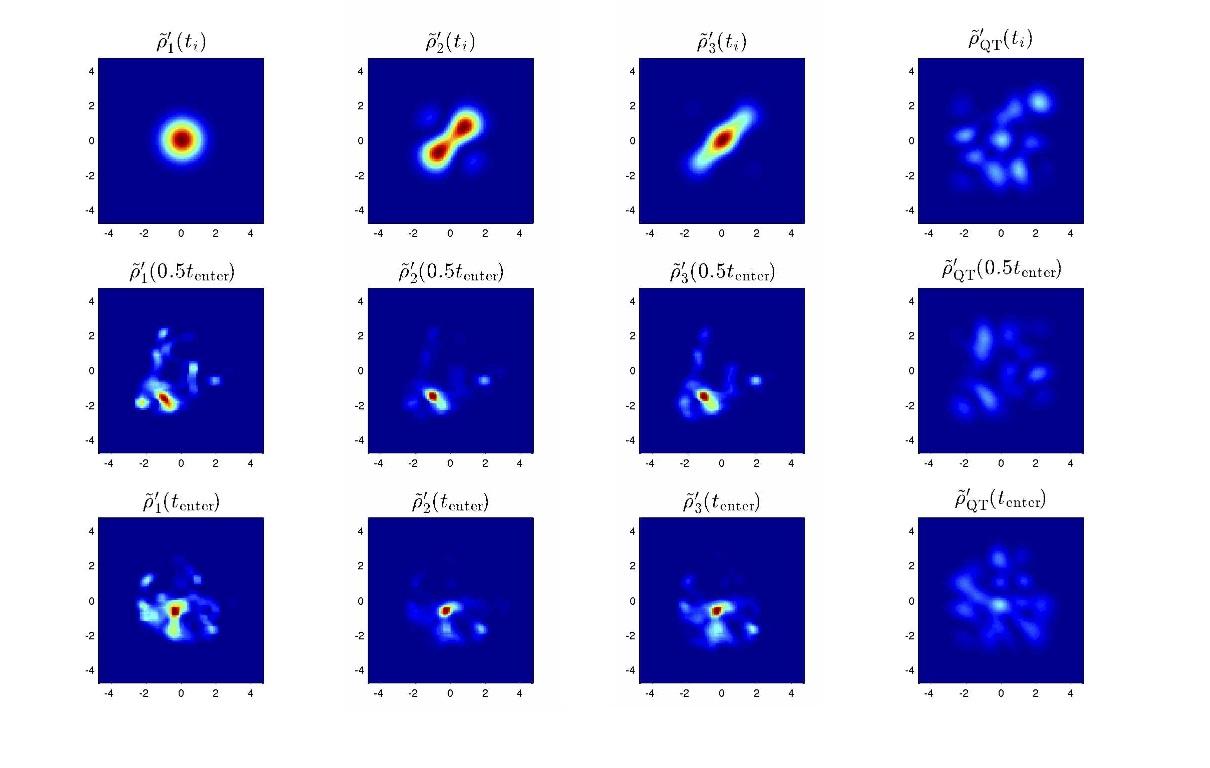}%
\caption{Relaxation suppression at super-Hubble wavelengths on expanding space
for different initial nonequilibria.}%
\end{center}
\end{figure}

\subsection{Robustness of the inverse-tangent{} deficit $\xi(k)$}

In ref. \cite{CV15} we studied the deficit function $\xi(k)$ for varying
numbers $M$ of modes, evolved over a fixed time interval $[t_{i}%
,t_{f}]=[10^{-4},10^{-2}]$ and with the fixed initial nonequilibrium density
(\ref{rho1i}). We found a good fit to the inverse-tangent function
(\ref{numksi}) -- with small ($\lesssim10\%$) oscillations around the curve --
in all the cases $M=4,6,9,12,16,25$ that were studied. Here we consider
varying the initial nonequilibrium density with fixed $M$ and evolving over
the same fixed time interval $[10^{-4},10^{-2}]$. We choose the intermediate
value $M=9$.

The simulation is carried out as described in detail in ref. \cite{CV15}. In
particular, we plot mixed-ensemble curves $\xi=\xi(k)$ obtained by averaging
the variances of the densities over six sets of randomly-chosen phases in the
initial wave function (\ref{psi_i}). However, in contrast to ref. \cite{CV15},
here we consider three different initial nonequilibrium distributions.

As shown in Figure 3, we find good fits to the inverse-tangent function
(\ref{numksi}) for all three of the initial densities (\ref{rho1i}%
)--(\ref{rho3i}) studied.%

\begin{figure}
[ptb]
\begin{center}
\includegraphics[width=\textwidth]
{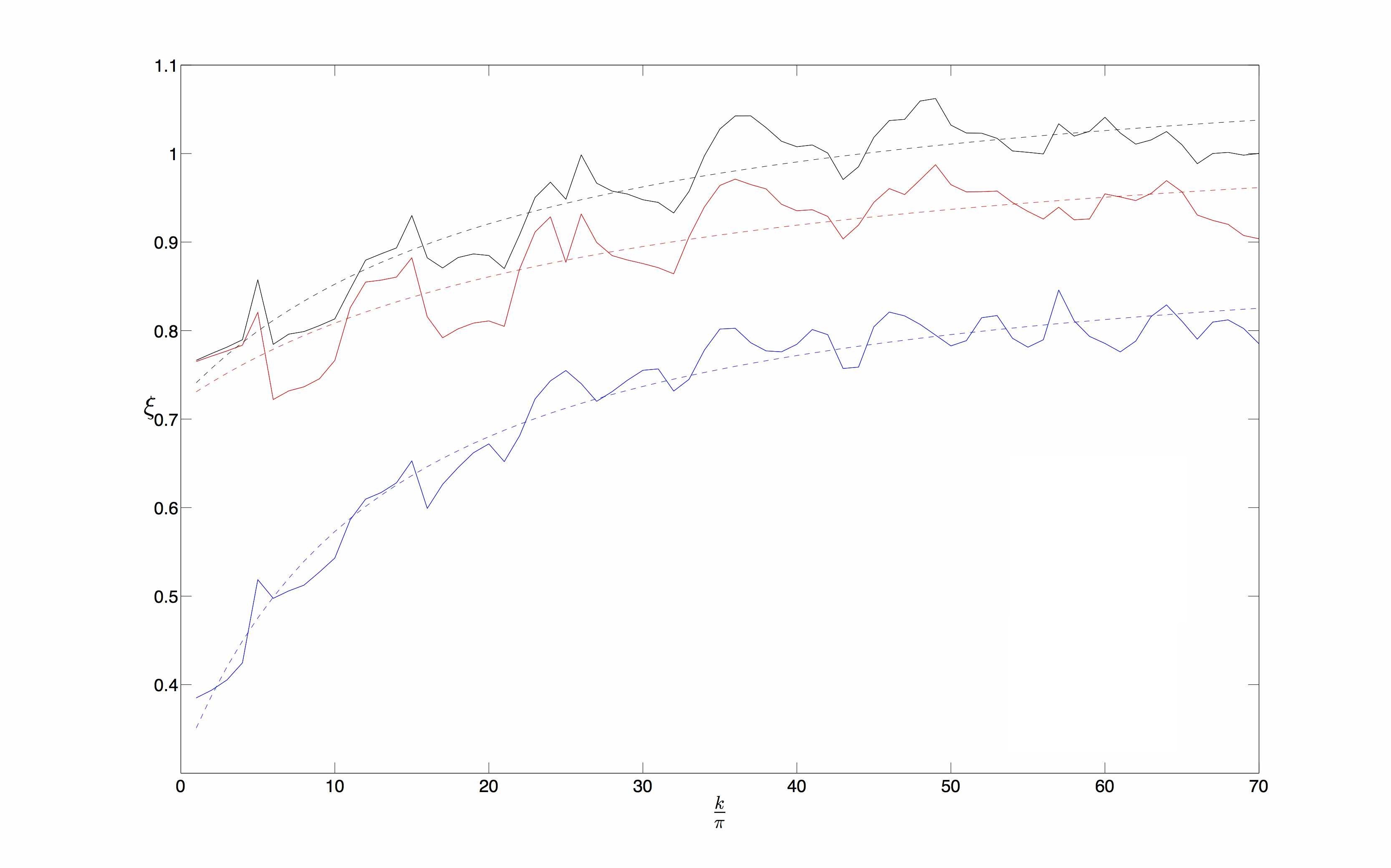}%
\caption{Mixed-ensemble curves $\xi(k)$ for the three initial nonequilibrium
densities (\ref{rho1i})--(\ref{rho3i}) ($\rho_{1}$ in blue, $\rho_{2}$ in red
and $\rho_{3}$ in black), with best fits to the inverse-tangent function
(\ref{numksi}).}%
\end{center}
\end{figure}

As shown in Table 1, the fitting coefficients $c_{1}$, $c_{2}$, $c_{3}$ depend
on the initial density. The first coefficient $c_{1}$ is almost the same for
the three initial densities but $c_{2}$ and $c_{3}$ vary considerably. Note
that for $\rho_{2}$ and $\rho_{3}$ we find $c_{3}>1$.

\begin{table}
\begin{center}%
\begin{tabular}
[c]{|c|c|c|c|}\hline
$\rho$ & $c_{1}$ & $c_{2}$ & $c_{3}$\\\hline
$\rho_{1}$ & 0.14 & 1.44 & 0.92\\
$\rho_{2}$ & 0.11 & 2.86 & 1.06\\
$\rho_{3}$ & 0.12 & 2.30 & 1.13\\\hline
\end{tabular}

\end{center}
\caption{Fitting coefficients $c_{1}$, $c_{2}$, $c_{3}$ for the initial
nonequilibrium densities $\rho_{1}$, $\rho_{2}$, $\rho_{3}$.}
\label{Table1}
\end{table}

\section{Radiation-dominated phase with a subsequent inflationary phase}

We now consider how the deficit function $\xi(k)$ will be affected by a period
of exponential (inflationary) expansion added to the end of a
radiation-dominated phase. We assume the fixed initial nonequilibrium density
(\ref{rho1i}) (with initial wavefunctions of the form (\ref{psi_i})) at time
$t_{i}$ at the beginning of the radiation-dominated period, which as before
lasts until a time $t_{f}$; we then add an inflationary period beginning at
$t_{f}$ and lasting until a time $t_{f}^{\prime}$. We shall calculate $\xi(k)$
for varying numbers $M$ of superposed energy states in the initial wavefunction.

To proceed, we must first find the retarded time function $t_{\mathrm{ret}%
}=t_{\mathrm{ret}}(t,k)$ for an exponential expansion. We may then add
appropriate retarded time intervals (corresponding to $[t_{f},t_{f}^{\prime}%
]$) to the evolution of the equivalent standard oscillator for modes of
wavenumber $k$, thereby obtaining the time evolution over the combined
radiation-dominated and inflationary eras. From the final densities -- again
simulated for a mixed ensemble with six sets of randomly-chosen initial phases
-- we are then able to calculate the mixed-ensemble deficit function $\xi(k)$.

\subsection{Retarded time for an inflationary phase}

The expression for the retarded time $t_{\mathrm{ret}}=t_{\mathrm{ret}}(t,k)$
arises \cite{CV13} by finding the solution for the time evolution $\psi
_{n_{r}}(q_{r},t)$ of the initial wave function $\psi_{n_{r}}(q_{r}%
,t_{i})=\Phi_{n_{r}}(q_{r})$ under a formal one-dimensional Schr\"{o}dinger
equation
\begin{equation}
i\frac{\partial\psi_{n_{r}}(q_{r},t)}{\partial t}=\left(  -\frac{1}%
{2m}\partial_{r}^{2}+\frac{1}{2}m\omega^{2}q_{r}^{2}\right)  \psi_{n_{r}%
}(q_{r},t)
\end{equation}
with Hamiltonian $\hat{H}_{r}(t)$ (equation (\ref{Ham})), where%
\begin{equation}
m\equiv a^{3}\ ,\ \ \ \omega\equiv k/a\ .
\end{equation}
The exact solution with $a\propto t^{1/2}$ was found in ref. \cite{CV13}. Here
we wish to find the solution with an inflating scale factor%
\begin{equation}
a(t)=a_{f}e^{H(t-t_{f})}\ , \label{a_inf}%
\end{equation}
where $t_{f}$ is the time at which the inflationary phase begins (while ending
at $t_{f}^{\prime}$) and $H$ is assumed to be constant.

As we saw in ref. \cite{CV13}, we first need to find two solutions $f_{1}$ and
$f_{2}$ of the classical equation%
\begin{equation}
\ddot{f}+\frac{\dot{m}}{m}\dot{f}+\omega^{2}f=0~. \label{osc_cl}%
\end{equation}
Then we define a function%
\begin{equation}
g_{-}=\gamma_{1}f_{1}^{2}+\gamma_{2}f_{1}f_{2}+\gamma_{3}f_{2}^{2}~,
\label{g-}%
\end{equation}
where $\gamma_{1}$, $\gamma_{2}$, $\gamma_{3}$ are constants (denoted $c_{1}$,
$c_{2}$, $c_{3}$ in ref. \cite{CV13}), and two other derived functions
\begin{equation}
g_{0}=-\frac{m}{2}\dot{g}_{-}\ ,\ \ \ \ g_{+}=m^{2}\omega^{2}g_{-}-m\dot
{g}_{0}\ . \label{g0+}%
\end{equation}
These functions must satisfy the initial conditions
\begin{equation}
g_{-}(t_{f})=\frac{1}{m_{f}}\ ,\ \ \ \ g_{0}(t_{f})=0\ ,\ \ \ \ g_{+}%
(t_{f})=m_{f}\omega_{f}^{2}\ . \label{ics}%
\end{equation}

The retarded time interval $\Delta t_{\mathrm{ret}}(k)$ corresponding to the
standard time interval $t_{f}^{\prime}-t_{f}$ is then given by the integral
\begin{equation}
\Delta t_{\mathrm{ret}}(k)=\int_{t_{f}}^{t_{f}^{\prime}}\frac{dt}%
{m(t)g_{-}(t)}~. \label{tret}%
\end{equation}
(See ref. \cite{CV13} for a detailed derivation.)

For the scale factor (\ref{a_inf}) we have $\dot{m}/m=3H$ and $\omega
=ke^{-H(t-t_{f})}/a_{f}$. Equation (\ref{osc_cl}) then becomes%
\begin{equation}
\ddot{f}+3H\dot{f}+\kappa^{2}e^{-2H(t-t_{f})}f=0~,
\end{equation}
where $\kappa=k/a_{f}$. We find the two solutions
\begin{align}
f_{1}  &  =\cos\left(  (\kappa/H)e^{-H(t-t_{f})}\right)  +(\kappa
/H)e^{-H(t-t_{f})}\sin\left(  (\kappa/H)e^{-H(t-t_{f})}\right)  \ ,\\
f_{2}  &  =\sin((\kappa/H)e^{-H(t-t_{f})})-(\kappa/H)e^{-H(t-t_{f})}%
\cos((\kappa/H)e^{-H(t-t_{f})})\ .
\end{align}

From (\ref{g-}) and (\ref{g0+}), together with the initial conditions
(\ref{ics}), we obtain the constants $\gamma_{1}$, $\gamma_{2}$, $\gamma_{3}%
$:
\begin{align}
\gamma_{1}  &  =\frac{1}{2k^{4}a_{f}}\left(  -H^{4}a_{f}^{2}\cos
(2k/Ha_{f})-2H^{3}ka_{f}\sin(2k/Ha_{f})+H^{2}(H^{2}a_{f}^{2}+2k^{2})\right)
\ ,\\
\gamma_{2}  &  =\frac{1}{k^{4}}\left(  2H^{3}k\cos(2k/Ha_{f})-Ha_{f}%
\sin(2k/Ha_{f})\right)  \ ,\\
\gamma_{3}  &  =\frac{1}{2k^{4}a_{f}}\left(  H^{4}a_{f}^{2}\cos(2k/Ha_{f}%
)+2H^{3}ka_{f}\sin(2k/Ha_{f})+H^{2}(H^{2}a_{f}^{2}+2k^{2})\right)  \ .
\end{align}
The expression (\ref{g-}) for $g_{-}$ is now fully determined.

For the purpose of obtaining the retarded time interval, the important
function is $mg_{-}$. Evaluating it explicitly, we find
\begin{align}
mg_{-}  &  =\left(  -\frac{1}{2\phi_{f}\phi^{3}}-\frac{2}{\phi^{2}}+\frac
{1}{2\phi_{f}\phi}\right)  \cos(2\phi-2\phi_{f})+\nonumber\\
&  \left(  \frac{1}{\phi^{3}}-\frac{1}{\phi_{f}\phi^{2}}-\frac{1}{\phi
}\right)  \sin(2\phi-2\phi_{f})+\nonumber\\
&  \left(  \frac{1}{2\phi_{f}\phi^{3}}+\frac{\phi_{f}}{\phi^{3}}+\frac
{1}{2\phi_{f}\phi}+\frac{\phi_{f}}{\phi}\right)  \ ,
\end{align}
where%
\begin{equation}
\phi\equiv(k/Ha_{f})e^{-H(t-t_{f})}\ .
\end{equation}

To evaluate the integral (\ref{tret}), we first change variables to $\phi$
where $d\phi=-H\phi dt$. The integrand becomes%
\begin{equation}
-\frac{2\phi_{f}}{H}\frac{\phi^{2}}{(-1-4\phi_{f}\phi+\phi^{2})\cos
(2\phi-2\phi_{f})+(2\phi_{f}-2\phi-2\phi_{f}\phi^{2})\sin(2\phi-2\phi
_{f})+(1+2\phi_{f}^{2}+\phi^{2}+2\phi_{f}^{2}\phi^{2})}\ .
\end{equation}
We then use $\cos2x=\cos^{2}x-\sin^{2}x$, $\sin2x=2\sin x\cos x$ and we
introduce two new functions
\begin{align}
u  &  =-\sin(\phi-\phi_{f})+\phi\cos(\phi-\phi_{f})\ ,\\
v  &  =-\frac{1}{\phi_{f}}\sin(\phi-\phi_{f})+\frac{\phi}{\phi_{f}}\cos
(\phi-\phi_{f})-\phi\sin(\phi-\phi_{f})-\cos(\phi-\phi_{f})~.
\end{align}
The integrand can then be written as
\begin{equation}
-\frac{1}{\phi_{f}H}\frac{(du/d\phi)v-u(dv/d\phi)}{u^{2}+v^{2}}\ .
\end{equation}
An indefinite integral with respect to $\phi$ is given by%
\begin{equation}
-\frac{1}{\phi_{f}H}\tan^{-1}(\frac{u}{v})\ .
\end{equation}
However, this is discontinuous at zeros of $v$. To proceed, the domain of
integration must be divided into subdomains and each time we cross a zero of
$v$ we must add $\pi a_{f}/k$ to the retarded time interval. (This requires
that we find the zeros of the function $v$, which is straightforward.)

\subsection{Robustness of the inverse-tangent{} deficit $\xi(k)$}

We now consider the effect of adding an inflationary phase. We take the
preceding radiation-dominated period to begin at $t_{i}=10^{-4}$ and end at
$t_{f}=10^{-2}$ just as before. We then take the inflationary phase to begin
at $t_{f}=10^{-2}$ and end at $t_{f}^{\prime}=t_{f}+0.05=0.06$, with an
inflationary Hubble constant $H=25$. The parameters have been chosen so as to
yield an interesting physical range and also to enable us to carry out the
numerical simulation in a reasonable amount of computer time. The chosen
values are not intended to have any particular cosmological
significance.\footnote{We use natural units with $\hslash=c=1$, so that time
has dimensions of an inverse mass. An initial time of $10^{-4}$ in our units
corresponds to $10^{-4}\hbar\sim10^{-37}\ \mathrm{s}$ in standard units.}

As shown in Figure 4, for these parameters the range of physical wavelengths
that we consider are mostly sub-Hubble at the beginning of inflation
($t=t_{f}$) and essentially all super-Hubble at the end of inflation
($t=t_{f}^{\prime}$). The simulation will then probe what happens when modes
exit the Hubble radius during inflation -- a key component of the standard
inflationary scenario.%

\begin{figure}
[ptb]
\begin{center}
\includegraphics[width=\textwidth]
{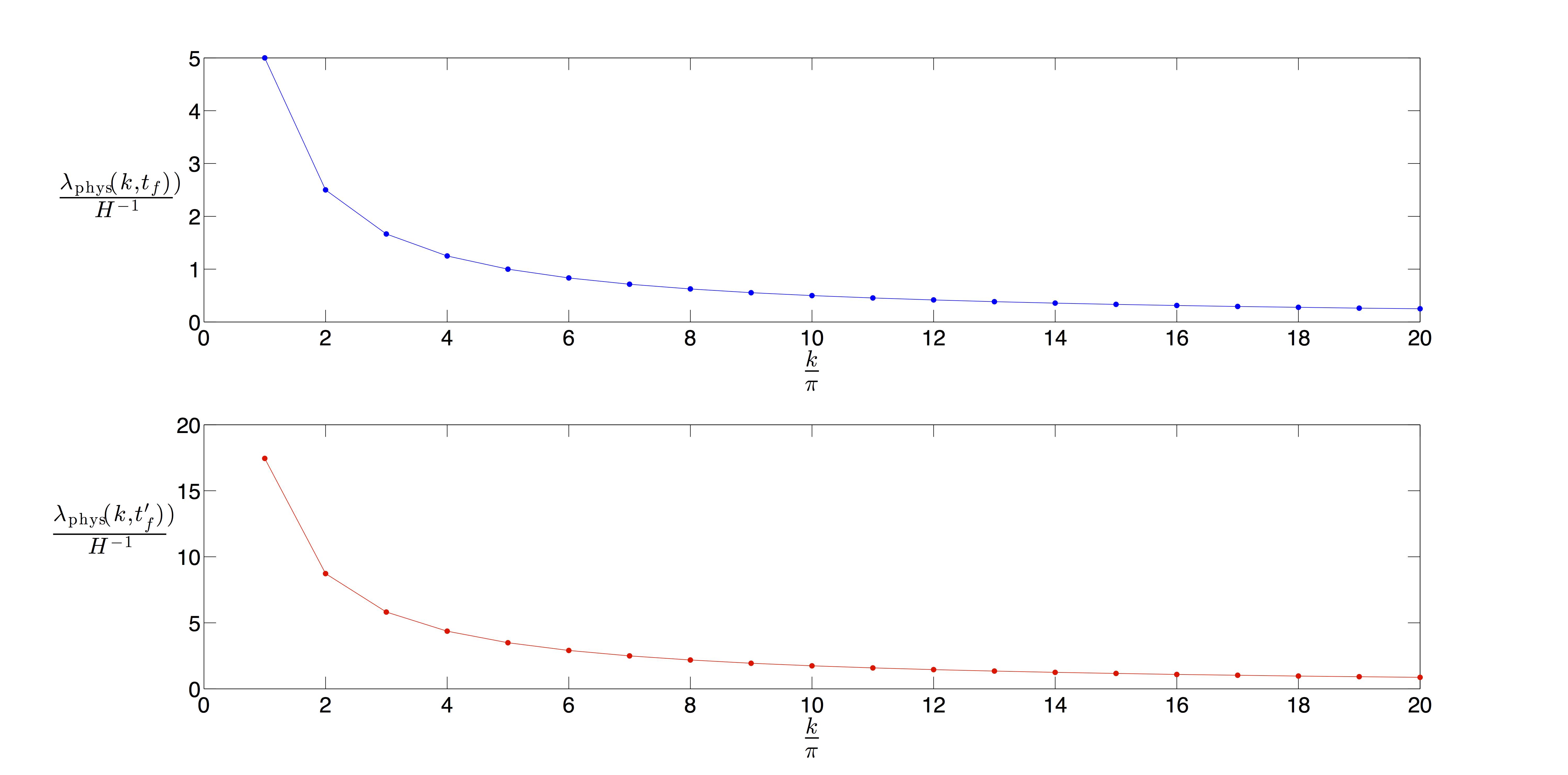}%
\caption{The top figure shows physical wavelengths as a function of $k$ at the
beginning of inflation ($t=t_{f}=10^{-2}$), in units of the Hubble radius. For
the range we consider, most of the modes are initially sub-Hubble. The bottom
figure shows physical wavelengths as a function of $k$ at the end of inflation
($t=t_{f}^{\prime}=0.06$), again in units of the Hubble radius. For the range
we consider, the final modes are essentially all super-Hubble.}%
\end{center}
\end{figure}

Furthermore, for this set of parameters the values of the retarded time
(plotted in Figure 5) are not so large as to prevent a numerical simulation in
an acceptable amount of computer time. It is noteworthy that as $k$ increases
the value of $\Delta t_{\mathrm{ret}}(k)$ quickly saturates.%

\begin{figure}
[ptb]
\begin{center}
\includegraphics[width=\textwidth]
{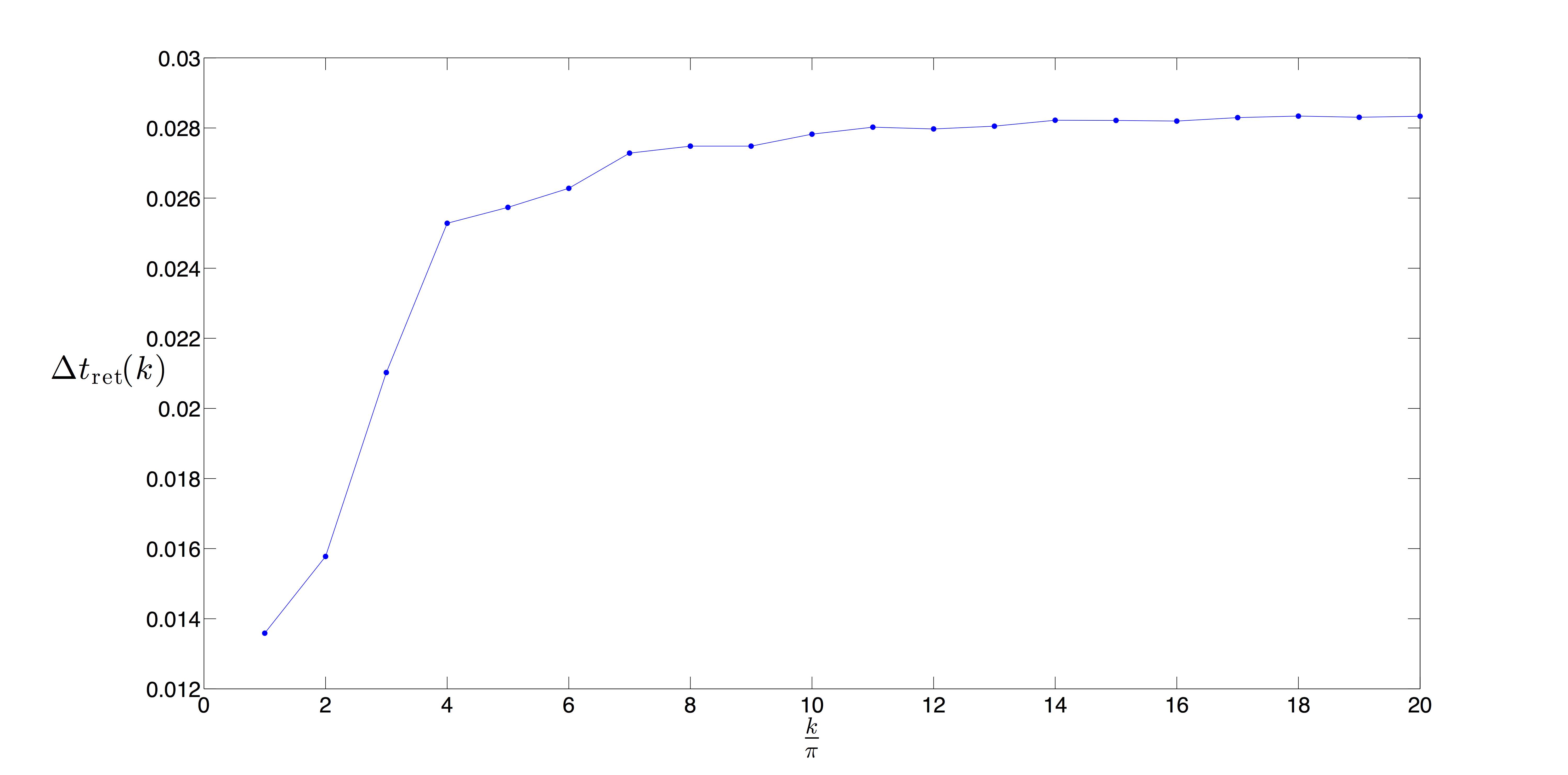}%
\caption{Retarded time interval $\Delta t_{\mathrm{ret}}(k)$ corresponding to
the lapse $t_{f}^{\prime}-t_{f}$ of standard cosmological time during the
inflationary phase, plotted as a function of wavenumber $k$.}%
\end{center}
\end{figure}

For this set of parameters, then, we numerically simulate the function
$\xi(k)$ for the combined radiation and inflationary eras, taking initial
wavefunctions (\ref{psi_i}) that are superpositions of $M$ energy states and
with the fixed initial nonequilibrium density (\ref{rho1i}). The simulation is
again carried out as described in ref. \cite{CV15}, plotting mixed-ensemble
curves $\xi=\xi(k)$ obtained by averaging the variances over six sets of
randomly-chosen initial phases, except that now the retarded time interval
(for the evolution of the equivalent oscillator) is increased by the value
$\Delta t_{\mathrm{ret}}(k)$ corresponding to the additional lapse
$t_{f}^{\prime}-t_{f}$ of standard cosmological time during the inflationary
phase. Our results for $M=4,6,9,12,16$ are shown in Figure 6. For each curve
in the figure we include a fit to the inverse-tangent function (\ref{numksi}).%

\begin{figure}
[ptb]
\begin{center}
\includegraphics[width=\textwidth]
{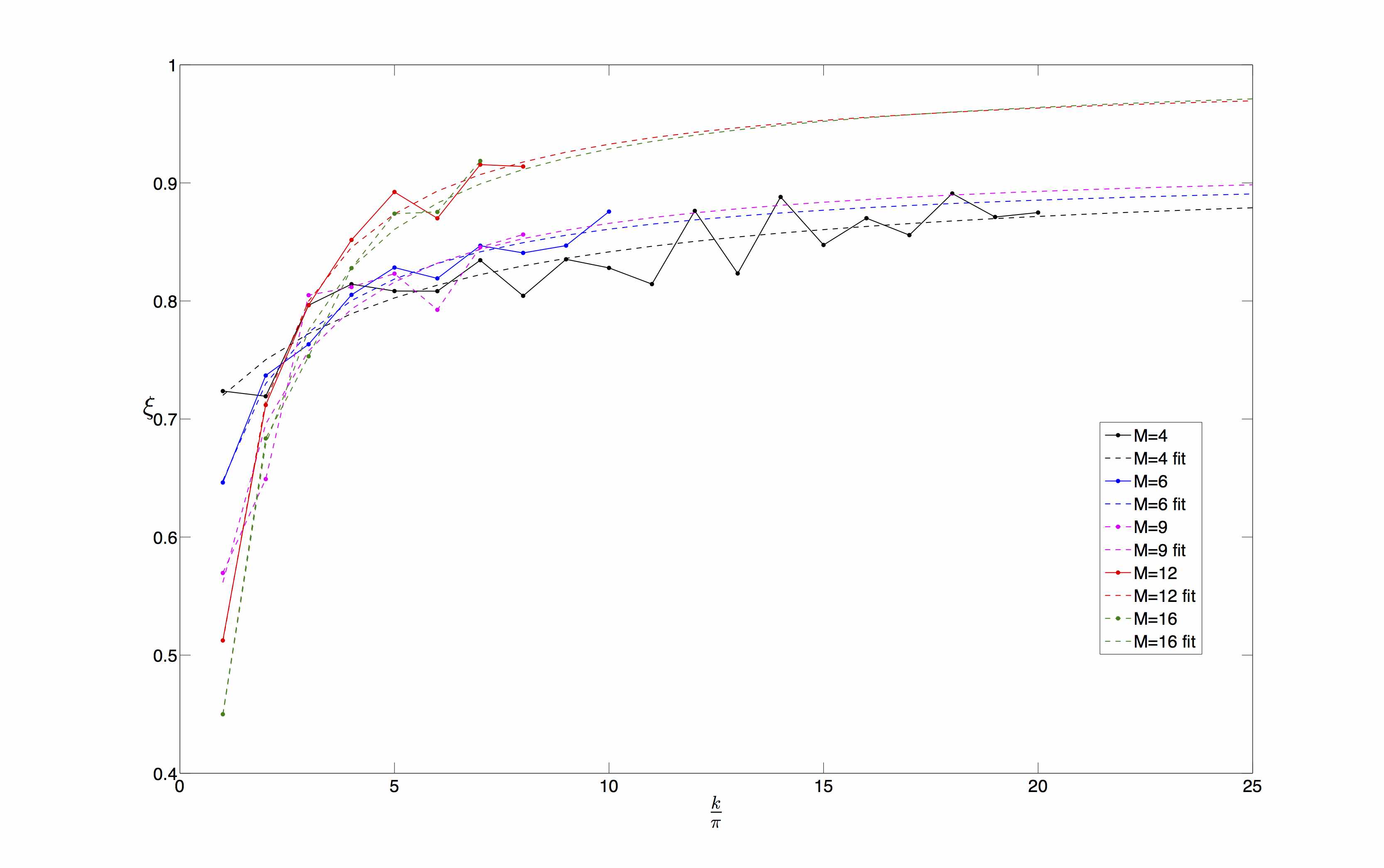}%
\caption{Mixed-ensemble curves $\xi(k)$ for the combined radiation-dominated
and inflationary eras, for $M=4,6,9,12,16$ modes and with the fixed initial
nonequilibrium density (\ref{rho1i}), with fits to the inverse-tangent
function (\ref{numksi}).}%
\end{center}
\end{figure}

For smaller $M$ we are able to integrate further in $k$-space since the
velocity field is less erratic for smaller $M$. In Figure 6 we show all of the
data that we were able to obtain to good accuracy in an acceptable amount of
computer time.

Despite the limited extent of our results in $k$-space for larger $M$, the
goodness of fit to the inverse-tangent function (\ref{numksi}) is clear
enough. Thus we may conclude that the inverse-tangent deficit (\ref{numksi})
is robustly valid under the addition of an inflationary phase.

\subsection{Dependence of the fitting coefficients on $M$}

Given the fits to the function (\ref{numksi}) for $M=4,6,9,12,16$, we may
study how the fitting coefficients $c_{1}$, $c_{2}$, $c_{3}$ vary with $M$.

Since we have generated data for all of the considered values of $M$ only up
to about $k=10\pi$, in order to compare like with like we consider it better
to omit the extra data at $k>10\pi$ for the case $M=4$. Thus, for $M=4$ we
truncate the curve at $k=10\pi$ before finding the fitting coefficients for
the inverse-tangent function (\ref{numksi}). With this understanding, the
resulting coefficients $c_{1}$, $c_{2}$, $c_{3}$ are shown in Table 2.

\begin{table}
\begin{center}%
\begin{tabular}
[c]{|c|c|c|c|}\hline
$M$ & $c_{1}$ & $c_{2}$ & $c_{3}$\\\hline
$4$ & $2.39$ & $4.01$ & $0.87$\\
$6$ & $1.72$ & $1.96$ & $0.91$\\
$9$ & $1.70$ & $0.96$ & $0.92$\\
$12$ & $1.58$ & $0.32$ & $0.99$\\
$16$ & $1.37$ & $0.25$ & $1.00$\\\hline
\end{tabular}
\end{center}
\caption{Fitting coefficients $c_{1}$, $c_{2}$, $c_{3}$ for varying
$M=4,6,9,12,16$.}
\label{Table2}
\end{table}

In Figure 7 we show how $c_{1}$ varies with $M$, together with a fitted curve%
\begin{equation}
c_{1}=1.51+8.55e^{-{0.57M}}{\ .} \label{c1vM}%
\end{equation}
This has the same functional form (with different coefficients) as the fit
found in our previous simulations for the radiation-dominated phase only
\cite{CV15}. While the fit is not quite as good as before, our results are
consistent with the previous (approximately exponential) decay of $c_{1}$ with
$M$.%

\begin{figure}
[ptb]
\begin{center}
\includegraphics[width=0.9\textwidth]
{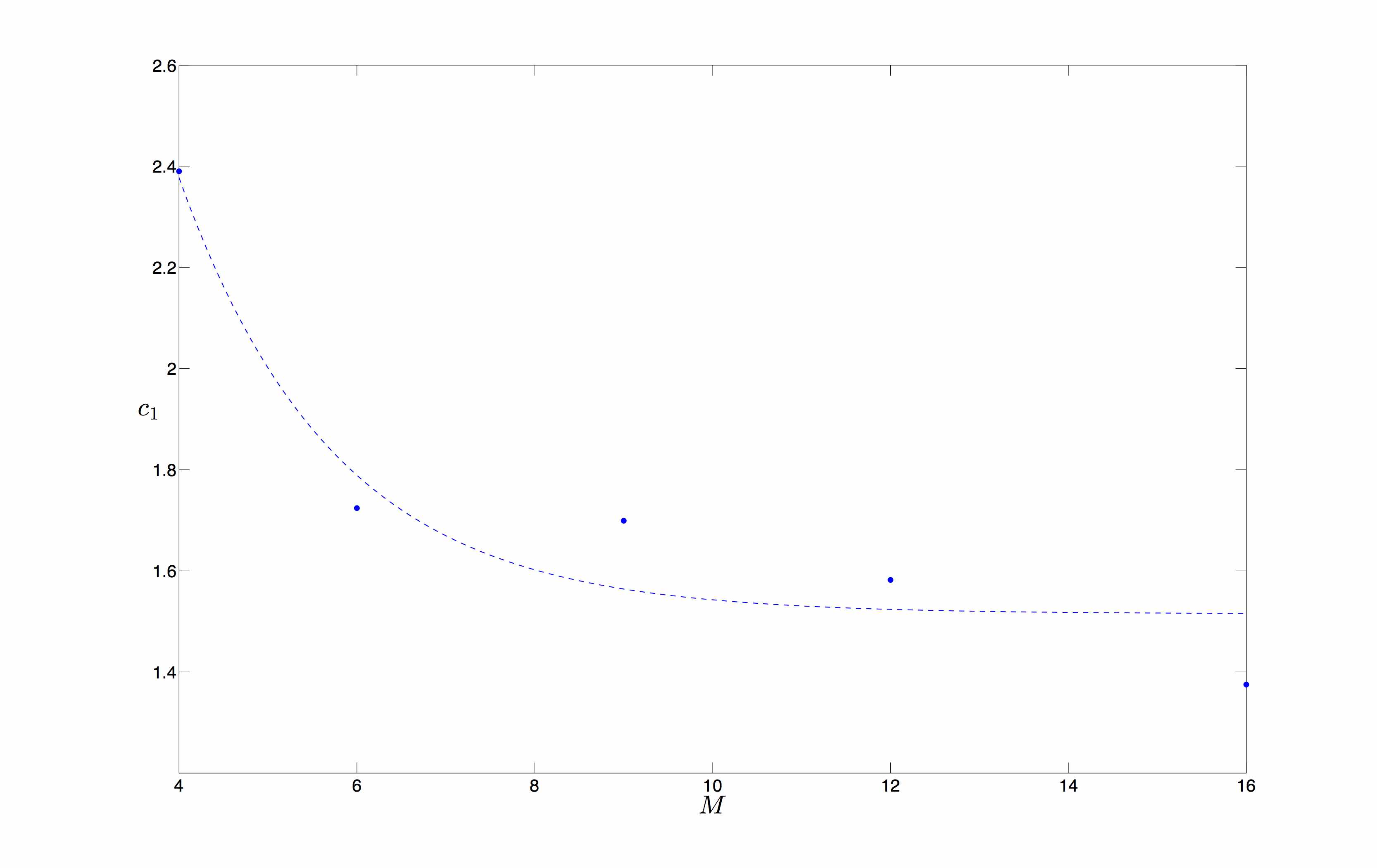}%
\caption{Plot of $c_{1}$ as a function of $M$, with a best-fit curve
(\ref{c1vM}).}%
\end{center}
\end{figure}

In Figure 8 we show how $c_{2}$ varies with $M$, together with a fitted curve%
\begin{equation}
c_{2}=0.20+16.11e^{{-0.36M}}\ . \label{c2vM}%
\end{equation}
Again, this has the same functional form as the fit found in our previous
simulations \cite{CV15} and now the goodness of fit is comparable to that
found previously.%

\begin{figure}
[ptb]
\begin{center}
\includegraphics[width=0.9\textwidth]
{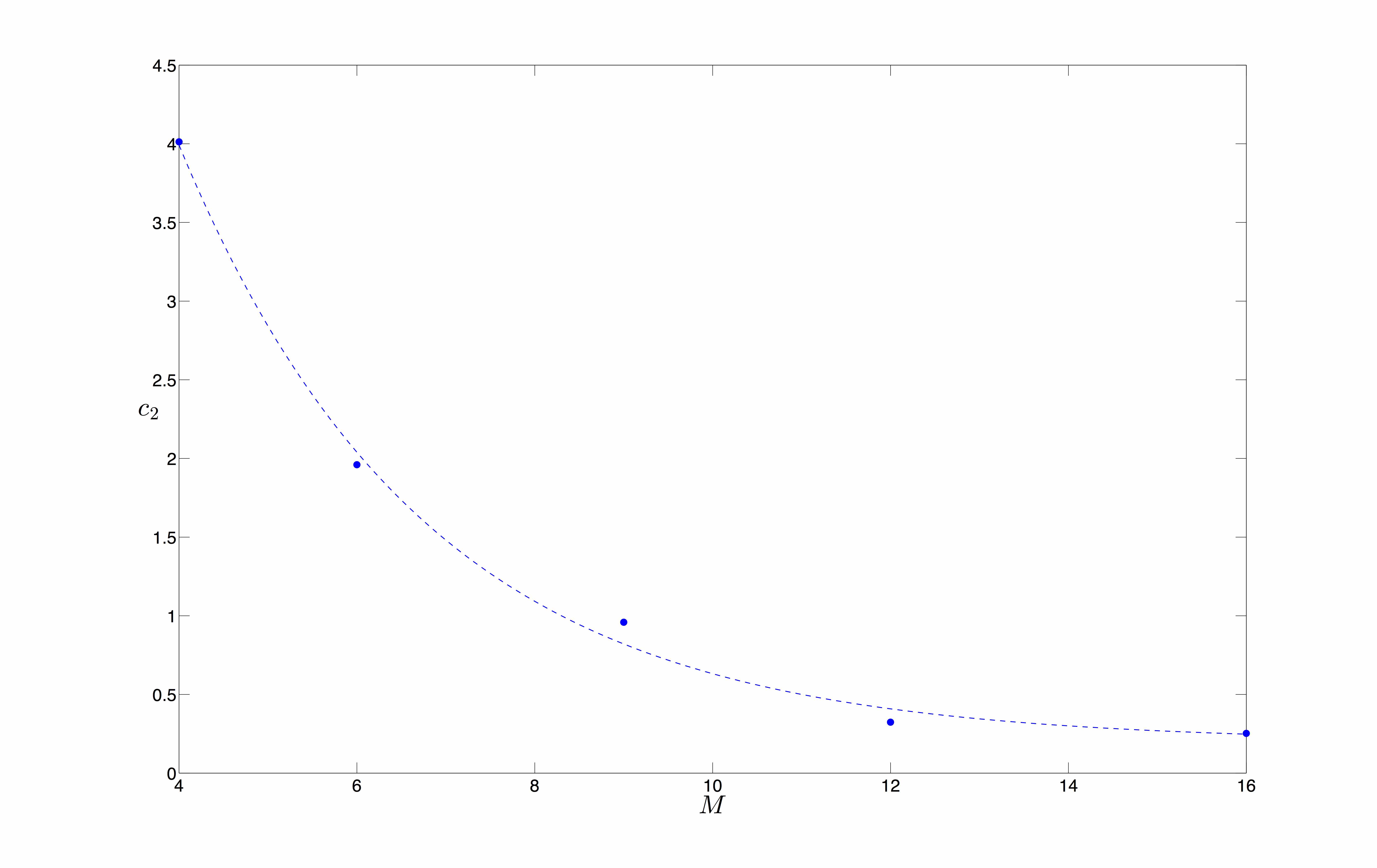}%
\caption{Plot of $c_{2}$ as a function of $M$, with a best-fit curve
(\ref{c2vM}).}%
\end{center}
\end{figure}

Finally, in Figure 9 we show how $c_{3}$ varies with $M$, together with a
fitted curve%
\begin{equation}
c_{3}=\tan^{-1}{(0.15M+2.11)}-\pi/2+1.22\ . \label{c3vM}%
\end{equation}
Once again this has the same functional form as the fit found previously
\cite{CV15} and with a fit that is not quite as good as before.%

\begin{figure}
[ptb]
\begin{center}
\includegraphics[width=0.9\textwidth]
{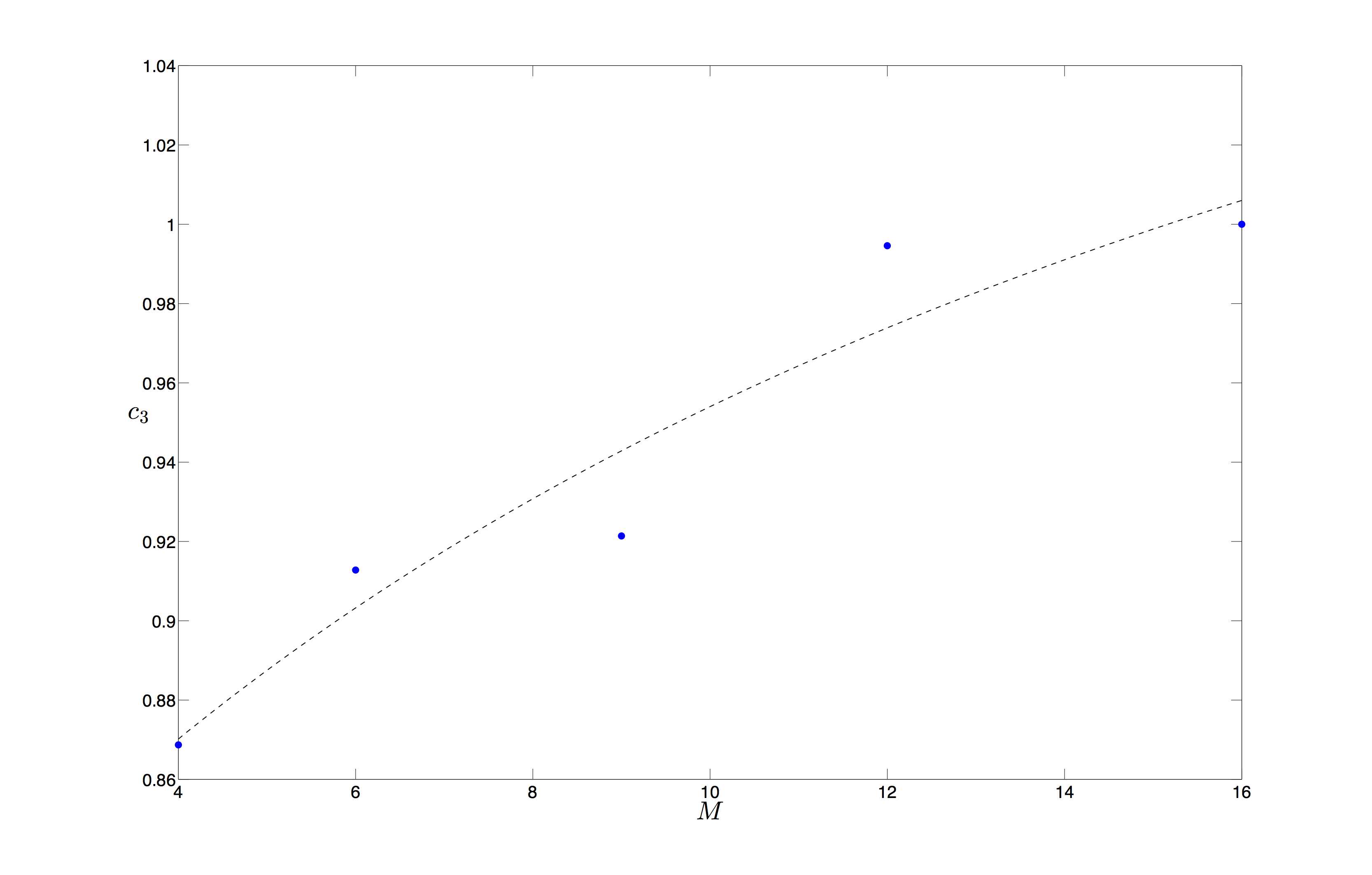}%
\caption{Plot of $c_{3}$ as a function of $M$, with a best-fit curve
(\ref{c3vM}).}%
\end{center}
\end{figure}

To a first approximation, our results for $c_{1}$, $c_{2}$, $c_{3}$ as
functions of $M$ are consistent with the results found in ref. \cite{CV15} for
the radiation-dominated phase only. Thus, not only is the inverse-tangent fit
(\ref{numksi}) robustly valid under the addition of an inflationary phase, the
functional dependence of the fitting coefficients on $M$ is also approximately maintained.

\section{Conclusion}

We have shown that our prediction of an inverse-tangent power deficit function
$\xi(k)$ -- arising from quantum relaxation on expanding space -- is robust
under changes in the initial nonequilibrium distribution as well as under the
addition of an inflationary period to the end of the radiation-dominated
phase. In both cases the simulated deficit $\xi(k)$ remains an inverse-tangent
function of $k$. Furthermore, with the inflationary phase the dependence of
the fitting parameters $c_{1}$, $c_{2}$, $c_{3}$ on the number $M$ of
superposed pre-inflationary energy states is found to be comparable to the
dependence found previously. These results suggest that, for the assumed broad
class of initial conditions, the inverse-tangent function (\ref{numksi}) is
likely to be a rather general signature of quantum relaxation in the early
universe. It remains to be seen how well this prediction compares with the
data \cite{PVV15}.

There is clearly a limit, however, to the robustness of our results against
changes in the initial nonequilibrium distribution. As an extreme example, if
a field mode had an initial delta-function distribution $\rho(q_{1}%
,q_{2},t_{i})=\delta(q_{1}-q_{1}(t_{i}))\delta(q_{2}-q_{2}(t_{i}))$ the
corresponding theoretical ensemble would contain just one trajectory beginning
at $q_{1}(t_{i})$, $q_{2}(t_{i})$ and ending at $q_{1}(t_{f})$, $q_{2}(t_{f}%
)$. The nonequilibrium distribution at later times would still be a
delta-function, $\rho(q_{1},q_{2},t_{f})=\delta(q_{1}-q_{1}(t_{f}%
))\delta(q_{2}-q_{2}(t_{f}))$. The final variance would vanish and we would
have $\xi=0$ for that mode. This could in principle happen for a range of
modes or even for all modes, resulting in no primordial power at all. Clearly,
for relaxation considered over a given time interval, if the initial
distribution is sufficiently narrow the final function $\xi(k)$ could differ
considerably from our inverse-tangent result (\ref{numksi}). In our
simulations we have taken initial states that depart fairly mildly from
quantum equilibrium, and for these we have found a final inverse-tangent
deficit over a range of initial conditions. In this sense our results are
robust. But if the changes in the initial state are too extreme, then with a
fixed time interval our inverse-tangent prediction will necessarily fail. On
the other hand, if the considered time interval is not fixed, then even for
very narrow initial nonequilibria one might obtain results resembling ours if
only the modes are evolved over a sufficiently long cosmological time. We
leave a study of this last point for future work.

Some assumptions certainly need to made about initial conditions, otherwise it
is impossible to make predictions. In our simulations we have always
considered initial distributions $\rho(q_{1},q_{2},t_{i})$ of width smaller
than the width of $|\psi(q_{1},q_{2},t_{i})|^{2}$. This may be justified
partly on heuristic grounds: if we regard quantum noise as having a dynamical
origin, it is natural to assume initial conditions such that the initial state
has less noise (or a smaller statistical spread) than a conventional quantum
state. We also assume that $\rho(q_{1},q_{2},t_{i})$ is appropriately smooth,
in the sense of not possessing any significant fine-grained microstructure --
an assumption that is required to obtain relaxation in any time-reversal
invariant theory \cite{AV92,AV96,AV01,VW05}. Our assumed initial conditions
have been guided by simplicity and heuristic arguments. Their ultimate
justification rests on how well our predictions compare with data.

We should also mention a further caveat. We have restricted ourselves to
initial quantum states (at the beginning of pre-inflation) of the form
(\ref{psi_i}). These are equally-weighted superpositions of energy eigenstates
with randomly-chosen phases. One might ask if our results are also robust
under changes to more general initial quantum states. In fact, all previous
studies of relaxation have assumed such equally-weighted superpositions. We
leave the study of more general superpositions for future work.

There is a remaining gap in our scenario. We do not have a proper treatment of
the transition from pre-inflation to inflation. In our simulations, we regard
the behaviour of the free scalar field $\phi$ as representative of the
behaviour of whatever fields may have been present during pre-inflation. We
have not specified the relation between $\phi$ and the inflaton field. Our
simulations yield a correction $\xi(k)$ to the power spectrum of $\phi$ and we
simply assume that the same correction appears in the inflationary spectrum
\cite{AV10,CV13,CV15}. Adding an inflationary period to our simulations of
relaxation for $\phi$ is a step towards filling the gap in our model. However,
a proper understanding requires a full field-theoretical model of the
transition from pre-inflation to inflation, which may involve symmetry
breaking and associated field redefinitions. We hope to develop such a model
in future work, and to run the appropriate numerical simulations required to
obtain the exact predicted deficit function $\xi(k)$.

\textbf{Acknowledgements}. We are grateful to Murray Daw for kindly providing
us with extra computational resources on the Clemson University Palmetto
Cluster. This research was supported in part by the John Templeton Foundation
and by Clemson University.

\end{document}